\documentclass[manuscript]{aastex}

\usepackage{emulateapj5}

\slugcomment{Accepted for publication in ApJ Letters}

\shorttitle{Activity at the Deuterium-Burning Mass Limit in Orion}
\shortauthors{Zapatero Osorio et al.}

\begin{document}

\title{Activity at the Deuterium-Burning Mass Limit in Orion}

\author{M.\,R. Zapatero Osorio\altaffilmark{1}}
\affil{Division of Geological and Planetary Sciences, California Institute of Technology MS\,150--21, Pasadena, CA 91125, U.S.A.}
\email{mosorio@gps.caltech.edu,mosorio@laeff.esa.es}

\author{V.\,J.\,S. B\'ejar}
\affil{Instituto de Astrof\'\i sica de Canarias, E-38200 La Laguna, Tenerife, Spain}
\email{vbejar@ll.iac.es}

\author{E.\,L. Mart\'\i n}
\affil{Institute of Astronomy, Univ. of Hawaii at Manoa, 2680 Woodlawn Drive, Honolulu, HI 96822, U.S.A.}
\email{ege@ifa.hawaii.edu}

\author{D. Barrado y Navascu\'es}
\affil{LAEFF-INTA, P.O$.$ 50727, E-28080 Madrid, Spain}
\email{barrado@laeff.esa.es}

\and

\author{R. Rebolo}
\affil{Instituto de Astrof\'\i sica de Canarias, E-38200 La Laguna, Tenerife, Spain, and Consejo Superior de Investigaciones Cient\'\i ficas, Madrid, Spain}
\email{rrl@ll.iac.es}

\altaffiltext{1}{Currently at: LAEFF-INTA, P.O$.$ 50727, E-28080 Madrid, Spain}

\begin{abstract}
  We report very intense and variable H$\alpha$ emission
  (pseudo-equivalent widths of $\sim$\,180, 410\,\AA) of S\,Ori\,55, a
  probable free-floating, M9-type substellar member of the young
  $\sigma$\,Orionis open star cluster.  After comparison with
  state-of-the-art evolutionary models, we infer that S\,Ori\,55 is
  near or below the cluster deuterium-burning mass borderline, which
  separates brown dwarfs and planetary-mass objects. We find its mass
  to be 0.008--0.015\,$M_{\odot}$ for ages between 1\,Myr and 8\,Myr,
  with $\sim$0.012\,$M_{\odot}$ the most likely value at the cluster
  age of 3\,Myr. The largest H$\alpha$ intensity reached the
  saturation level of log\,$L_{\rm H\alpha}$/$L_{\rm bol}$\,=\,--3. We
  discuss several possible scenarios for such a strong emission.  We
  also show that $\sigma$\,Orionis M and L dwarfs have in general more
  H$\alpha$ emission than their older field spectral
  counterparts. This could be due to a decline in the strength of the
  magnetic field with age in brown dwarfs and isolated planetary-mass
  objects, or to a likely mass accretion from disks in the very young
  $\sigma$\,Orionis substellar members.
\end{abstract}


\section{Introduction}
Since the discovery of substellar objects (unable to fuse hydrogen
stably in their interiors and with masses below 0.072\,$M_{\odot}$,
Kumar \cite{kumar63}; Chabrier et al$.$ \cite{chabrier00a}), many
efforts have been devoted to their characterization (see Basri
\cite{basri00} for a review). Recently, Muench et al$.$
\cite{muench01} found evidence for the presence of disks around brown
dwarfs (BDs) of the Trapezium cluster.  Muzerolle et al$.$
\cite{muzerolle00} and Mart\'\i n et al$.$ \cite{martin01a} reported
on disk accretion in T\,Tauri objects with masses near the substellar
borderline. All these works may suggest that ``circumstellar'' disks
are common among very young BDs. These objects generally show
H$\alpha$ in emission, which seems to be variable. To the best of our
knowledge, only a few BDs with masses larger than 0.06\,$M_{\odot}$
have been observed flaring (Rutledge et al$.$ \cite{rutledge00}; Basri
\& Mart\'\i n \cite{basri99}).

Here, we report on the detection of strong H$\alpha$ emission in
S\,Ori\,55 (S\,Ori\,J053725.9--023432), which was previously
identified as a cool, very low mass substellar member of the young
$\sigma$\,Orionis open cluster (Zapatero Osorio et al$.$
\cite{osorio00}; Barrado y Navascu\'es et al$.$ \cite{barrado01}).
Table~\ref{data} summarizes available spectrophotometric data for this
object. 

\section{Observations and Results \label{observations}}
We acquired low-resolution optical spectra of S\,Ori\,55 with the red
module of the Low-Resolution Imaging Spectrograph (Oke et al$.$
\cite{oke95}) at the 10-m Keck\,I telescope (Mauna Kea Observatory) on
2001 January 31. We used the 2048$\times$2048 pixel SITE detector
(0.25\arcsec\,pix$^{-1}$), the 150\,lines\,mm$^{-1}$ grating blazed at
750.0\,nm, the OG570 filter for blocking the light blueward of
570.0\,nm, and a 1.5\arcsec-width slit, which provides a wavelength
coverage of 600--1030\,nm, a spectral nominal dispersion of
4.0\,\AA\,pix$^{-1}$ and a final resolution of
24\,\AA~($R$\,$\sim$\,350). The slit was rotated to be at a
parallactic angle and minimize refraction losses. Two individual
exposures of 2400\,s each were obtained at different positions
separated by 6\arcsec~along the entrance slit. Weather conditions were
spectrophotometric, with seeing around 1.5\arcsec. Raw images were
processed with standard techniques that include bias subtraction and
flat-fielding within {\sc iraf\footnote{IRAF is distributed by
National Optical Astronomy Observatory, which is operated by the
Association of Universities for Research in Astronomy, Inc., under
contract with the National Science Foundation.}}. Nodded images were
subtracted to remove Earth's atmospheric contribution. A
full-wavelength solution (the $rms$ of the fifth-order polynomial fit
was 1\,\AA) was achieved by calibrating sky emission lines as in
Osterbrock et al$.$ \cite{osterbrock96}. To complete data reduction,
we corrected for the instrumental response using data of the
spectrophotometric standard star G\,191--B2B obtained on the same
night and with the same instrumental configuration.

\begin{figure}
\plotone{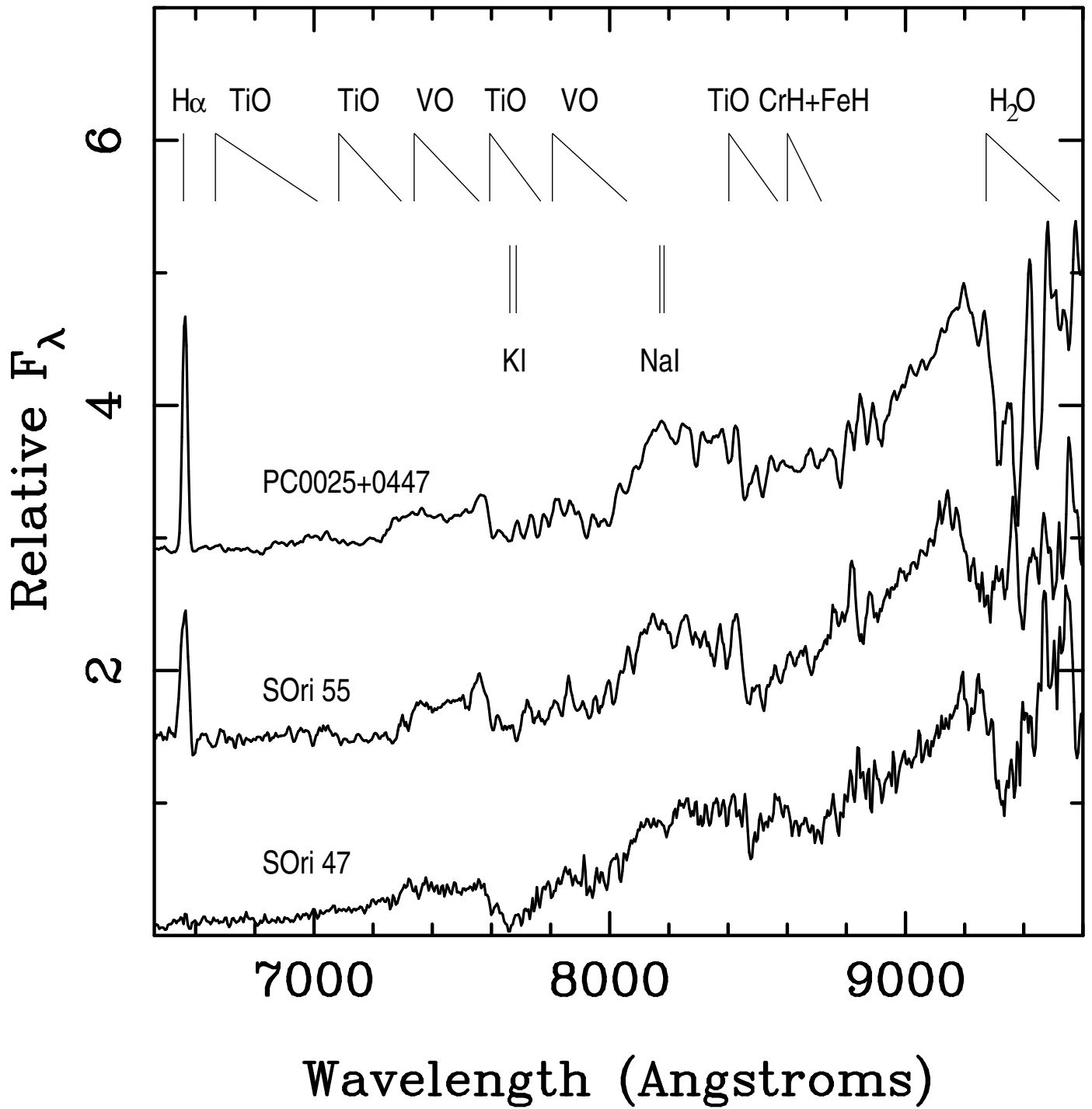}
\caption{Keck low-resolution averaged spectrum of S\,Ori\,55
  (M9). S\,Ori\,47 (L1.5, another substellar member of the
  $\sigma$\,Orionis cluster) and the field dwarf PC\,0025+0447 (M9.5)
  are also shown for comparison (spectra taken from Mart\'\i n et al$.$
  \cite{martin99}). A boxcar smoothing of 5 pixels has been applied to
  the data of S\,Ori\,55 and PC\,0025+0047. All spectra are normalized
  to the counts at $\sim$8150\,\AA. Upper spectra are shifted by 1.4
  units each for clarity. Main molecular and atomic features are
  indicated as in Kirkpatrick et al$.$ \cite{kirk99}.  \label{lr}}
\end{figure}

The resultant combined spectrum of S\,Ori\,55 is depicted in
Figure~\ref{lr} (S/N$\sim$10 at 750\,nm, and S/N$\sim$15 at
905\,nm). The spectra of the $\sigma$\,Orionis substellar member
S\,Ori\,47 (L1.5, Zapatero Osorio et al$.$ \cite{osorio99}) and the
field dwarf PC\,0025+0447 (M9.5, Kirkpatrick, Henry, \& Simons
\cite{kirk95}) are also shown for comparison. We derived the spectral
type of our target by matching its observed spectrum to data of
standards, which were previously obtained with similar
instrumentation. We also used the PC3 index of Mart\'\i n, Rebolo, \&
Zapatero Osorio \cite{martin96}. We measured an M9 spectral class with
an uncertainty of half a subclass, in full agreement with the previous
assignement of Barrado y Navascu\'es et al$.$ \cite{barrado01} and
with the observed $IJK$ colors.  Interestingly, S\,Ori\,55 displays
typical low-gravity features (Allard et al$.$ \cite{allard01};
Mart\'\i n et al$.$ \cite{martin96}; Luhman, Liebert, \& Rieke
\cite{luhman97}; B\'ejar, Zapatero Osorio, \& Rebolo \cite{bejar99}),
such as the absence of Na\,{\sc i} $\lambda\lambda$818.3,\,819.5\,nm,
and K\,{\sc i} $\lambda\lambda$762.1,\,766.7\,nm atomic lines at the
resolution of our data, and stronger molecular absorptions of VO and
TiO compared to field spectral counterparts. Our measured upper limits
to the line pseudo-equivalent widths (pEWs, relative to the local
observed pseudo-continuum formed by molecular absorptions) are given
in Table~\ref{data}. These values are significantly smaller than those
of older objects in the Pleiades and the field (Mart\'\i n et al$.$
\cite{martin96}; Zapatero Osorio et al$.$ \cite{osorio97}).

\begin{figure}
\plotone{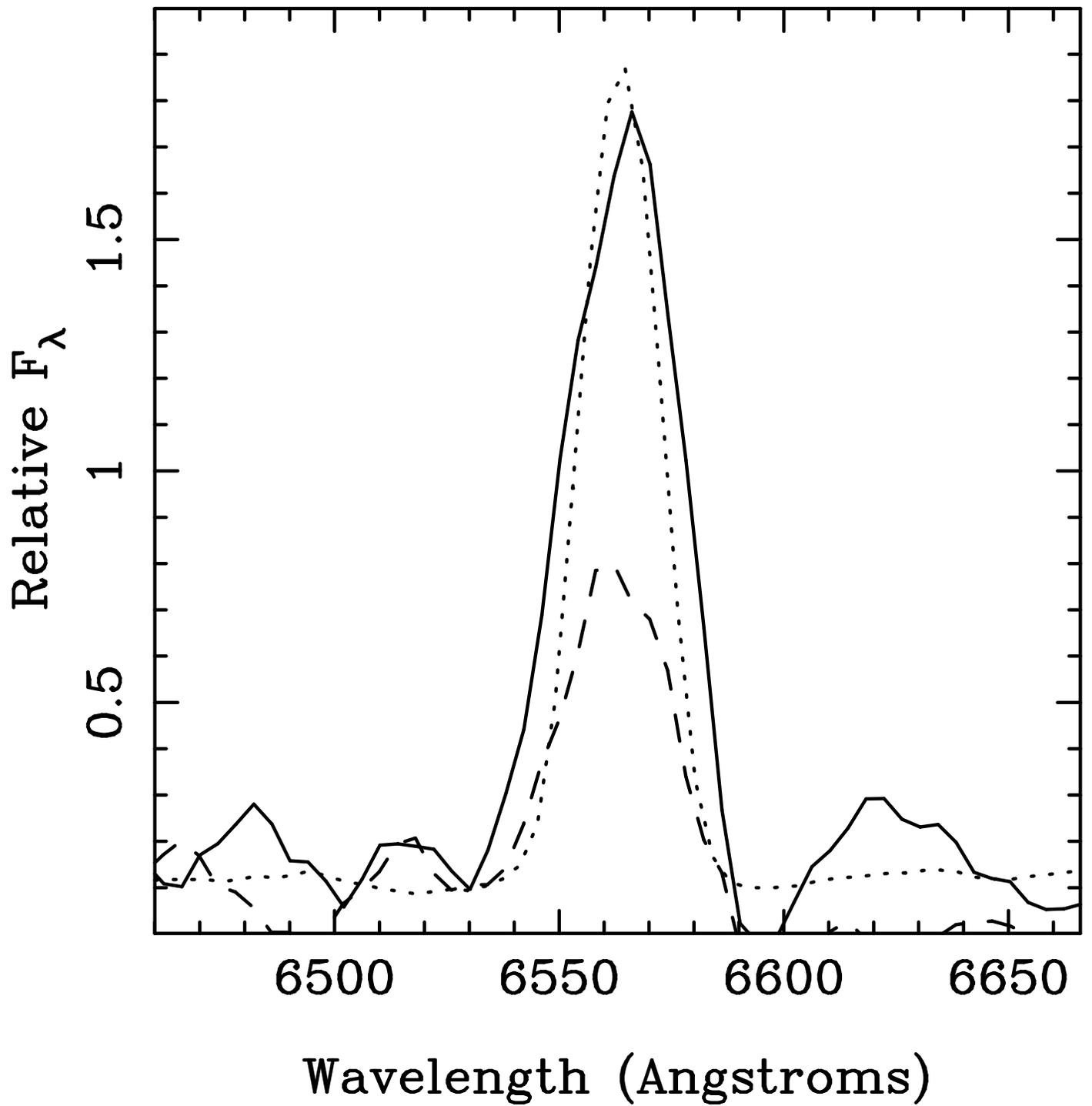}
\caption{H$\alpha$ profiles of S\,Ori\,55 (first 
  spectrum---dashed line; second spectrum---solid line).  
  PC\,0025+0447 (dotted line) is included for comparison (same 
  spectral resolution). Observed
  fluxes are normalized as in Fig.~\ref{lr}.  \label{ha}}
\end{figure}

The Keck optical data of S\,Ori\,55 show a rather strong H$\alpha$
line, which is seen in emission and with a different intensity between
the two closely spaced spectra. A close-up region around H$\alpha$ is
displayed in Figure~\ref{ha}. S\,Ori\,55 appears very active; however
we found the same spectral type as Barrado y Navascu\'es et al$.$
\cite{barrado01}, suggesting that no significant continuum veiling of
the visible photospheric features is present in the Keck spectra.
Measuring H$\alpha$ pEWs is not easy because the number of counts at
the line pedestal is small (50--140\,counts). In Table~\ref{data} we
provide heliocentric Julian dates and our best pEW measurements
obtained via direct integration of the line profile adopting the
pseudo-continuum level immediately adjacent to the line (around
$\lambda$652.5 and $\lambda$664.0\,nm). The error bars were obtained
after integrating over a reasonable range of possible continua. We are
confident that the emission is larger in the second Keck spectrum than
in the first one, which was taken 43 minutes earlier. Barrado y
Navascu\'es et al$.$ \cite{barrado01} found a less intense emission
(pEW\,=\,5\,\AA) in their VLT optical data taken $\sim$1 month before
the Keck observations. This result and our observations indicate that
S\,Ori\,55 has a variable H$\alpha$ emission.

Spectroscopy, and optical and near-IR photometry are all consistent
with S\,Ori\,55's being a member of the $\sigma$\,Orionis cluster. But
late-M stars are also common in the galactic disk, so we have
considered the probability that S\,Ori\,55 might simply be an
interloper field M9 dwarf in the direction toward the cluster (lying
at a distance between 118\,pc and 197\,pc in a volume of roughly
142\,pc$^3$). For this exercise we have used three-band photometry and
the absolute magnitudes of field stars (Leggett et al$.$
\cite{leggett02}).  Very recent all-sky surveys show that the space
density of M8--M9.5 dwarfs per cubic parsec is 0.0045\,$\pm$\,0.0008
(Gizis et al$.$ \cite{gizis00}). In our survey of the
$\sigma$\,Orionis cluster (Zapatero Osorio et al$.$ \cite{osorio00}),
we found five candidates (including S\,Ori\,55) with M9--M9.5 spectral
types in an interval of roughly 1\,mag (Barrado y Navascu\'es et al$.$
\cite{barrado01}). The probability of each candidate being a true
member is in the range 87--93\%. The finding of such a large number of
cool M dwarfs in our survey can be understood in terms of the much
higher object density in the $\sigma$\,Orionis cluster than in the
field.

\section{Mass Determination \label{mass}}
The very likely membership of S\,Ori\,55 in the $\sigma$\,Orionis
cluster allows us to derive its mass by comparison with
state-of-the-art evolutionary models. Its effective temperature
($T_{\rm eff}$) can be inferred from the spectral type and the
optical-infrared colors. The calibration of Basri et al$.$
\cite{basri00al} gives $T_{\rm eff}$\,=\,2370\,K for an M9 dwarf,
while Luhman's \cite{luhman99} scale, which is intermediate between
dwarfs and giants, provides $T_{\rm eff}$\,=\,2550\,K. Recently,
B\'ejar \cite{bejar01} has derived a temperature calibration by
matching observed low-resolution optical spectra of cluster M-type BDs
to a spectral synthesis computed for the dusty, log\,$g$\,=\,3.5 model
atmospheres of Allard et al$.$ \cite{allard01} (see Pavlenko, Zapatero
Osorio, \& Rebolo \cite{pav00}). B\'ejar \cite{bejar01} found $T_{\rm
eff}$\,=\,2500\,K for an M8.5 cluster member. Because S\,Ori\,55 has a
surface gravity around log\,$g$\,=\,4.0, on the basis of theoretical
models, its $T_{\rm eff}$ could be in the range 2370--2550\,K. Less
uncertainties exist in the calculation of the bolometric
luminosity. Using the photometry of S\,Ori\,55 and the bolometric
corrections of Monet et al$.$ \cite{monet92}, Reid et al$.$
\cite{reid01}, and Leggett et al$.$ \cite{leggett02}, we derived
log\,$L/L_{\odot}$\,=\,--3.08\,$\pm$\,0.25. The uncertainty is mainly
due to the error in the Hipparcos distance (Perryman et al$.$
\cite{perryman97}) to the massive central star of the cluster.

\begin{figure}
\plotone{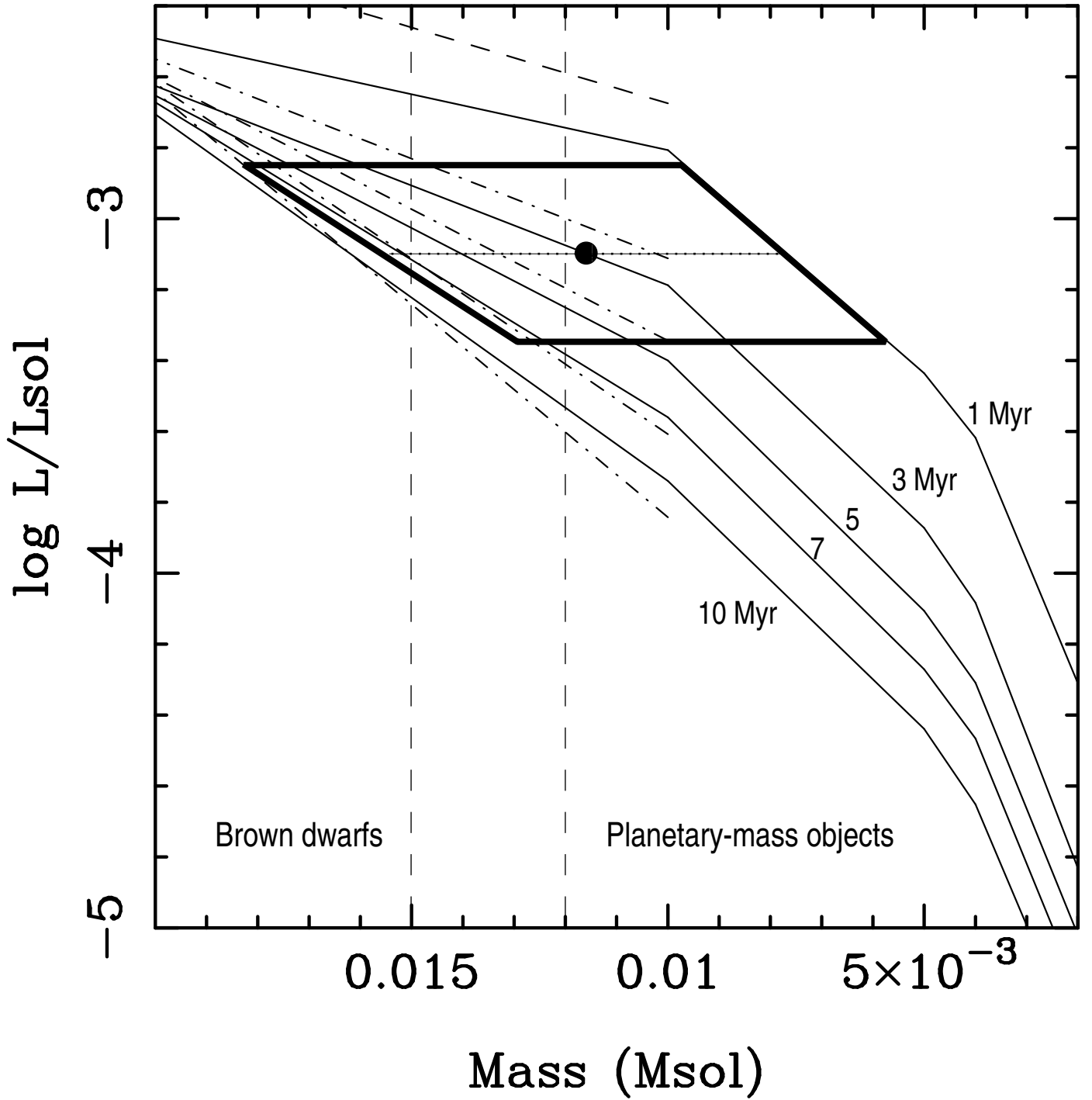}
\caption{Mass of S\,Ori\,55. Models (1--10\,Myr) of Chabrier et al$.$ 
  \cite{chabrier00a} and
  Burrows et al$.$ \cite{burrows97} are plotted with solid and
  dash-dotted lines, respectively. Vertical dashed lines indicate the
  deuterium-burning mass limit (borderline between brown dwarfs and
  planetary-mass objects; Saumon et al$.$ \cite{saumon96}). The age of
  S\,Ori\,55 is believed to be between 1\,Myr and 8\,Myr with the most
  likely value at $\sim$3\,Myr (filled circle). The mass of S\,Ori\,55 
  is 0.012\,$\pm$\,0.004\,$M_{\odot}$ for log\,$L/L_{\odot}$\,=\,--3.08 
  (horizontal dotted line). \label{sori55}}
\end{figure}

Figure~\ref{sori55} illustrates the substellar mass-luminosity
relationships of Chabrier et al$.$ \cite{chabrier00a} and Burrows et
al$.$ \cite{burrows97}. These two sets of evolutionary models deal
with dusty atmospheres differently. While the former models assume
that dust particles are formed in the atmosphere and remain there
impacting the output energy distribution, the latter models treat dust
as if condensed below the photosphere. S\,Ori\,55 may be coeval with
other cluster members; thus its mass can be easily deduced from the
figure. The age of the $\sigma$\,Orionis association is discussed to
some extent in Zapatero Osorio et al$.$ \cite{osorio02}.  These
authors concluded that the most likely cluster age is 2--4\,Myr,
although ages as young as 1\,Myr and as old as 8\,Myr cannot be
discarded. Further discussion of other cluster properties can be found
in B\'ejar et al$.$ \cite{bejar01al} and Walter, Wolk, \& Sherry
\cite{walter98}. In Figure~\ref{sori55} the vertical lines between
0.012\,$M_{\odot}$ (Chabrier et al$.$ \cite{chabrier00b}) and
0.015\,$M_{\odot}$ (Burrows, Hubbard, \& Lunine \cite{burrows89})
denote the deuterium-burning mass limit, which separates BDs and
planetary-mass objects. Values as high as 0.018\,$M_{\odot}$ can be
found in the literature (D'Antona \& Mazzitelli \cite{dantona94}). The
luminosity of S\,Ori\,55, its error bar, and possible ages are
indicated with the region enclosed by the thick line. The filled
circle stands for the most likely values of luminosity and age. Using
the Chabrier et al$.$ \cite{chabrier00a} models, the mass of
S\,Ori\,55 is estimated at 0.012\,$\pm$\,0.004\,$M_{\odot}$, i.e., very
close to the deuterium-burning mass threshold. The tracks of Burrows
et al$.$ \cite{burrows97} yield smaller masses by $\sim$10\%.
S\,Ori\,55 is defining the cluster frontier between BDs and
planetary-mass objects. It could be a planetary-mass cluster member if
it were younger than $\sim$3\,Myr.

\section{Discussion and Final Remarks \label{final}}
This is the first time that very large H$\alpha$ emission has been
detected in a low mass substellar object. Its origin remains unknown.
S\,Ori\,55 may simply have undergone a chromospheric flare due to
magnetic activity. This is supported by the rapid time variability of
the line and the lack of a constant, significant emission.
Furthermore, no additional continuum appears to be veiling the
photospheric spectral features. Many dwarfs at the bottom of the M
class experience strong and weak flares (Gizis et al$.$
\cite{gizis00}; Mart\'\i n \& Ardila \cite{martin01b}). Based on our
data, we tentatively estimate that the flaring recurrence of S\,Ori\,55
is about 33--66\%, which contrasts with the 7\%~value of the field
coolest M dwarfs (Gizis et al$.$ \cite{gizis00}).

Similarly intense H$\alpha$ lines have been detected in just a few
field objects of related $T_{\rm eff}$. Liebert et al$.$
\cite{liebert99} observed a strong flare in 2MASSW\,J0149090+295613
(M9.5) with H$\alpha$ pEWs of 200--300\,\AA, and many other emission
lines of He\,{\sc i}, O\,{\sc i}, K\,{\sc i} and Ca\,{\sc ii}. Mould
et al$.$ \cite{mould94} and Mart\'\i n, Basri, \& Zapatero Osorio
\cite{martin99} reported variable and persistent H$\alpha$ emission in
PC\,0025+0447 (M9.5) with pEWs between 100 and 400\,\AA.  Liebert et
al$.$ interpreted the behavior of the 2MASS object as that of a very
cool M-type flare star, and they pointed out that this flare activity
differs from that of PC\,0025+0447. Albeit with differences, the
optical spectra of PC\,0025+0447 and S\,Ori\,55 seem more alike.
Burgasser et al$.$ \cite{burgasser00} argued that the former object
could be an interacting binary. S\,Ori\,55 might also be double with
one of the components losing mass to the other. Our optical spectra do
not suggest the presence of a warm component or a significant
variation of the continuum over a month. ROSAT X-ray data provide an
upper limit of log\,($L_X/L_{\rm bol}$)\,$\le$\,--0.9 to the X-ray
emission of S\,Ori\,55 (Mokler \cite{mokler02}). We note, however,
that PC\,0025+0447 does not show significant X-ray flux (Neuh\"auser
et al$.$ \cite{neuhaeuser99}).

The strong H$\alpha$ emission of S\,Ori\,55 may have its origin in the
object formation processes. Haisch, Lada, \& Lada \cite{haisch01} found
that the presence of dusty disks surrounding stellar members of
clusters as young as the $\sigma$\,Orionis cluster is significantly
high (50\%--65\%). There are also clear indications of accretion
events among the low mass star population of $\sigma$\,Orionis
(Zapatero Osorio et al$.$ \cite{osorio02}). However, accreting
T\,Tauri stars show continuous strong emission, whereas S\,Ori\,55
appears to have episodes of low and high activity. It could be that
mass infall from protoplanetary disks onto central substellar-mass
objects turns out to be unstable due to nonuniform disks and variable
magnetic fields. Sporadic accretion events may also indicate the end
of the accreting activity in BDs and planetary-mass objects. The mass
infall rate seems to be rather small in objects around the substellar
limit (Muzerolle et al$.$ \cite{muzerolle00}). S\,Ori\,55 does not
show an infrared excess in the $K$ band, which is consistent with a
very low accretion rate.  With current data we cannot rule out any
possible scenario (flare activity, binarity, mass accretion).

\begin{figure}
\plotone{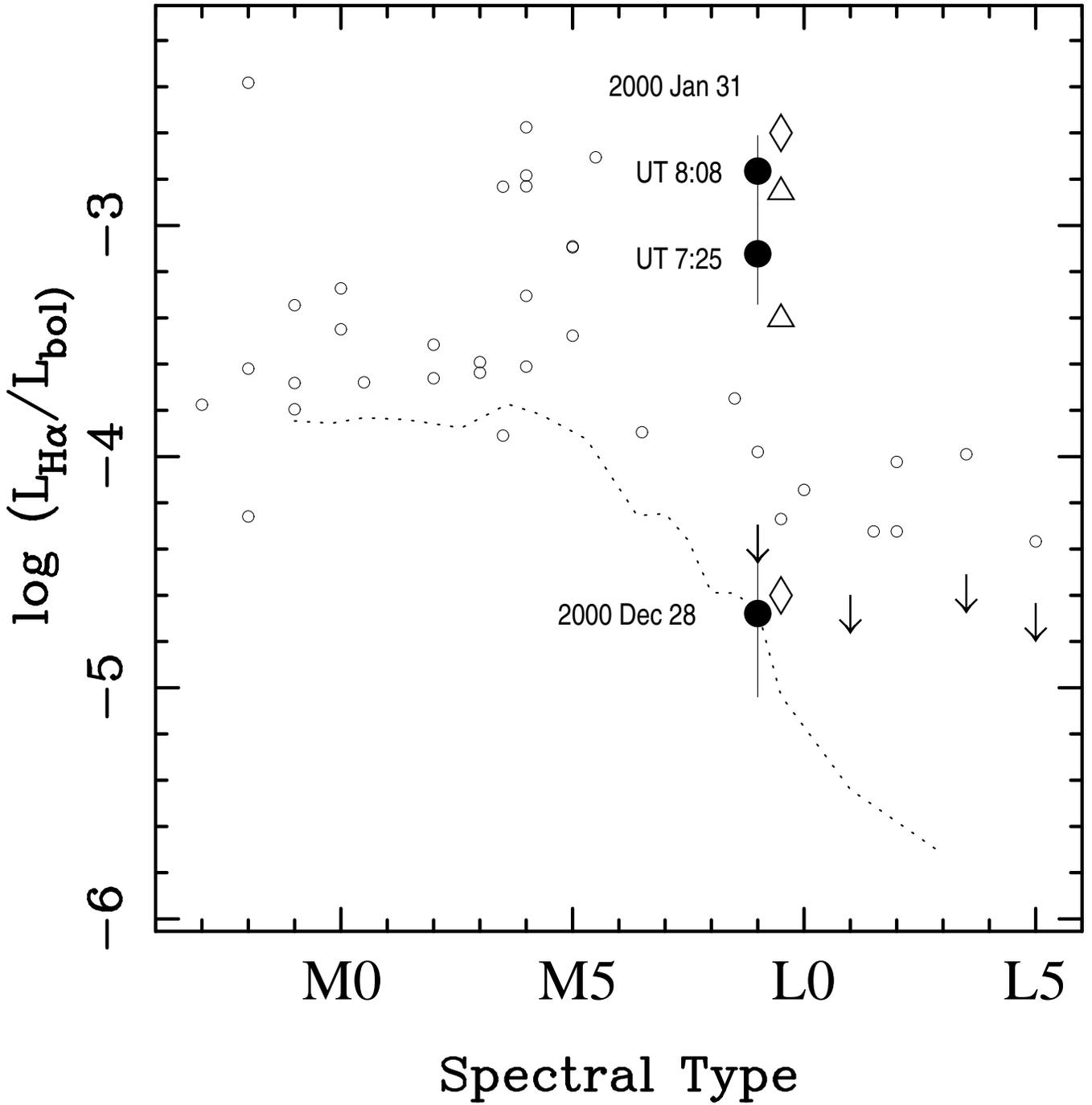}
\caption{Ratio of H$\alpha$ luminosity of the object to its 
  bolometric luminosity as a function of spectral type.
  $\sigma$\,Orionis members are plotted with open circles, and
  S\,Ori\,55 is indicated with filled circles. The
  star--brown dwarf boundary in $\sigma$\,Orionis takes place at
  around the M5 spectral type, and the brown dwarf--planetary-mass
  frontier at around the M9 class. The averaged locus of field stars is
  delineated by the dotted line. Ratios of PC\,0025+0447 (triangles,
  minimum and maximum H$\alpha$ emissions found in the literature) and
  2MASSW\,J0149090+295613 (diamonds, quiescent and flaring stages) are
  also plotted. \label{habol}}
\end{figure}

We have compared the H$\alpha$ emission to the bolometric luminosity
($L_{\rm bol}$) of $\sigma$\,Orionis low mass members from the late-K
spectral type (stellar regime) down to mid-L classes (planetary-mass
domain). H$\alpha$ fluxes ($L_{\rm H\alpha}$) have been calculated as
in Herbst \& Miller \cite{herbst89}, and bolometric corrections have
been taken from various sources in the literature (Monet et al$.$
\cite{monet92}; Leggett et al$.$ \cite{leggett02}).  Our results are
portrayed in Figure~\ref{habol}. For comparison purposes, we have also
included the averaged values of field stars with similar spectral
types (individual star data published by Hawley, Gizis, \& Reid
\cite{hawley96}, and Gizis et al$.$ \cite{gizis00}). The three
observations of S\,Ori\,55 are clearly indicated. The ratio $L_{\rm
H\alpha}$/$L_{\rm bol}$ of cluster members with warm spectral types
down to mid- and late-M classes appears to be rather dispersed around
log\,$L_{\rm H\alpha}$/$L_{\rm bol}$\,=\,--3.61\,dex (Zapatero Osorio
et al$.$ \cite{osorio02}). Cooler types (corresponding to
planetary-mass objects) display a less intense ratio by about
0.6\,dex, which implies a decrease in the activity level (less intense
magnetic fields). On average, $\sigma$\,Orionis members, and
remarkably those in the L classes, show higher H$\alpha$ emissions
than their older, field spectral counterparts. This feature could be
explained partly as a decline in the strength of the chromospheric and
flare activity with age and/or mass accretion from surrounding disks
in the young objects. This would indicate that magnetospheric mass
infall could extend even beyond the deuterium-burning mass threshold,
i.e., into the planetary-mass domain.

\acknowledgments We are grateful to I$.$ Baraffe and A$.$ Burrows for
providing us computer-ready files of their models. We also thank S$.$
Kulkarni for his assistance in collecting data with the Keck
telescope, and Louise Good for correcting the English language. This
paper is based on observations made with the W.\,M.  Keck Observatory,
which is operated as a scientific partnership among the California
Institute of Technology, the University of California and the National
Aeronautics and Space Administration (the Observatory was made
possible by the generous financial support of the W.\,M. Keck
Foundation).

\clearpage

\begin{deluxetable}{ccccccccccc}
\tabletypesize{\scriptsize}
\tablecaption{Data for S\,Ori\,55 (S\,Ori\,J053725.9--023432). \label{data}}
\tablewidth{0pt}
\tablehead{
  \colhead{$I$} & \colhead{$I-J$} & \colhead{$I-K$} & \colhead{SpT} &
  \colhead{$T_{\rm eff}$} & \colhead{log\,$L/L_{\odot}$} &
  \colhead{$M/M_{\odot}$} &
  \colhead{HJD\tablenotemark{a}} & \colhead{H$\alpha$\tablenotemark{b}}
  & \colhead{Na\,{\sc i}\tablenotemark{b}} 
  & \colhead{K\,{\sc i}\tablenotemark{b}}
}
\startdata 
21.32$\pm$0.03 & 3.10$\pm$0.07 & 4.32$\pm$0.10 & M9$\pm$0.5 & 
  2370--2550 & --3.08$\pm$0.25 & 0.012$\pm$0.004 &  6.7211 & 5$\pm$5  & & \\
&               &               &            &            &   
             &                 & 41.8093 & 180$\pm$80  & $\le$2 & $\le$16 \\
&               &               &            &            & 
             &                 & 41.8390 & 410$\pm$100 & $\le$2 & $\le$16 \\
\enddata

\tablenotetext{a}{HJD -- 2451900 (beginning of the exposures).}
\tablenotetext{b}{Pseudo-equivalent widths in \AA~(see text).}

\tablecomments{Photometric magnitudes taken from Zapatero Osorio et
  al$.$ \cite{osorio00}. $T_{\rm eff}$ is given in K. The first H$\alpha$ 
  pEW is taken from Barrado y Navascu\'es et al$.$ \cite{barrado01}.
  }

\end{deluxetable}

\clearpage

\clearpage

\clearpage

\clearpage

\end{document}